\documentclass[]{JHEP}
\usepackage{epsfig}

\newcommand{\leqsim}{\,\raisebox{-0.6ex}{$\buildrel < \over \sim$}\,}
\newcommand{\geqsim}{\,\raisebox{-0.6ex}{$\buildrel > \over \sim$}\,}
\newcommand{\be}{\begin{equation}}
\newcommand{\ee}{\end{equation}}
\newcommand{\ba}{\begin{eqnarray}}
\newcommand{\ea}{\end{eqnarray}}

\newcommand{\ie}{\mbox{\em i.e.~}}

\newcommand{\cf}{\mbox{\em c.f.~}}
\newcommand{\nn}{\nonumber}
\newcommand{\dif}{\mbox{d}}
\def\gev{\,{\rm GeV}}

\preprint{CERN-TH 99-289  
\\ DAMTP-1999-118 \\
\hepph{9909448}}

\title{Quasi-Fixed Points and Charge and Colour Breaking in 
Low Scale Models}

\author{S.A. Abel \\ Theory Division, Cern 1211, Geneva 23, Switzerland}
\author{B.C. Allanach \\ DAMTP, University of Cambridge, Wilberforce Rd, Cambridge, CB3 0WA,
United Kingdom}


\abstract{
We show that the current LEP2 lower bound upon the minimal supersymmetric
standard model (MSSM)
lightest Higgs mass rules out quasi-fixed scenarios for 
string scales between $10^6$ and $10^{11}$ GeV unless the heaviest stop mass
is more than 2 TeV.
We consider the implications of the low string scale for 
charge and colour breaking (CCB) bounds in the MSSM, and 
demonstrate that CCB bounds from $F$ and $D$-flat directions are
significantly weakened.
For scales less than $10^{10}$ GeV
these bounds become merely that degenerate scalar
mass squared values are positive at the string scale. 
}

\begin{document}

\section{Introduction}

For many years, string and unification scales were thought 
to be high ($\geqsim 10^{16}$
GeV). The perturbative heterotic formulation of string theory had the
fundamental string
scale $\Lambda_s \sim O(10^{17})$ GeV close to $M_{Planck} \sim 10^{19}$
GeV because of its constrained description of the gravitational interaction.
The grand unification (GUT) scale was around $\Lambda_{GUT} \sim 10^{16}$ GeV,
motivated by the apparent convergence of the gauge couplings when 
they were evolved to this value.
Recently however, attention has turned to models that have
lower string  and/or unification scales~\cite{dvali,
dienes,benakli,powerlaw,kobayashi,iban} and this has raised some 
interesting questions to do with renormalisation group evolution of parameters. 

The most immediate is of course whether gauge or Yukawa unification 
is still possible or even necessary with a lower string scale. 
One example that achieves gauge unification at the string scale~\cite{dienes} 
has the couplings experience power law 
`running'~\cite{dienes,powerlaw,kobayashi} above a compactification
scale due to the presence of additional Kaluza-Klein modes. 
A Kaluza-Klein spectrum with the same ratios of gauge beta functions as 
those in the MSSM leads to a logarithmic running up to the 
compactification scale with rapid power law unification taking place 
very rapidly thereafter~\cite{dienes}. 
An example that does not achieve gauge unification is `mirage'
unification~\cite{iban}.
In mirage unification the gauge couplings at the string scale receive moduli 
dependent corrections that behave as if there were continued logarithmic
running above the string scale up to unification at
the usual $\Lambda_{GUT}$. `Mirage unification' refers to this fictitious 
unification\footnote{Note that although there are problems with the particular
string realisation of mirage unification in ref.~\cite{iban}, the idea may be
realisable in other models and remains an interesting possibility.}.

A particularly attractive choice for the string scale (albeit 
one that is not immediately accessible to experiment) is 
$\Lambda_s \sim 10^{11}$ GeV~\cite{benakli}.
In this case the hierarchy
between the weak scale and the Planck scale arises 
without unnaturally small ratios of fundamental scales.
It was also noted in the first reference of~\cite{benakli} 
that $\Lambda_s \sim 10^{11}$ GeV gives neutrino masses of the right 
order. We return to this model below and refer to it as the 
Weak-Planck (WP) model.

In this paper we consider two other related issues in the 
Minimal Supersymmetric Standard Model (MSSM),
\be
W_{MSSM}=h_U Q H_2 U^c + h_D Q H_1 D^c + h_E L H_1 E^c+\mu H_1 H_2,
\ee 
with a low string scale.
The first concerns the top quark Quasi-Fixed Point (QFP).
The QFP is characterised by a focusing of some MSSM parameters to particular
ratios as the renormalisation scale $\Lambda$ is decreased towards the 
top quark mass, $m_t$~\cite{quasi,steveandme,haber}. Formally it is defined to
be 
the point in parameter space where there is a Landau pole in the top Yukawa
coupling $h_t$ at 
the string or GUT scale (whichever is the lower). 
In practice however this focusing behaviour can occur for a 
large but finite
$h_t(\Lambda_s)$, still treat-able by perturbation theory.
The coupling $h_t$ focusses to some value at $m_t$ independent
of $h_t(\Lambda_s)$ provided it is large enough. In low scale models, 
with their foreshortened logarithmic running, one naturally expects 
this behaviour to be very different. 
If the pole is at $\Lambda_s < \Lambda_{GUT}$, we expect the quasi-fixed
value of the top Yukawa at $m_t$ to
be larger than for the usual GUT scale unification.
Conversely, for a given value of top mass and $\tan\beta $ at the weak scale 
the model will be further from the QFP for $\Lambda_s < \Lambda_{GUT}$. 
We shall determine the QFP prediction for $h_t(m_t)$, on which experimental
constraints from LEP2 can be brought to bear 
in order to empirically constrain
$\Lambda_s$ assuming the QFP scenario. In particular, 
we consider the empirically derived lower bound upon the lightest CP-even
MSSM Higgs mass, which in
the canonical GUT scenarios has been shown
to be a strong restriction upon the QFP scenario~\cite{haber}.

The second issue we consider is the possibility of minima 
that break charge and colour lying along $F$ and $D$ flat 
directions~\cite{steveandme,ccb1,dilaton,us0,us,casas,dedes}. 
The constraints found by requiring that there be 
no such (CCB) minima are 
dependent on the distance from the QFP\@.
They are most severe at the QFP itself~\cite{steveandme,us0,us}
and indeed, in the usual MSSM at the QFP, CCB constraints 
exclude half the parameter space. 
With a lower string scale it seems likely 
that such constraints will generally be less restrictive
for two reasons. First, a given point in (weak-scale) parameter space 
will be further from the QFP as noted above.
Second, the CCB minima are generated radiatively when the 
mass-squared parameter for $H_2$ becomes negative. When there is 
a lower string scale there is less `room' for a minimum to form 
at vacuum expectation values (VEVs) much greater than the weak 
scale. (More specifically, there are positive mass-squared 
contributions to the potential along the flat direction that become  
dominant at lower VEVs.) We shall demonstrate that this is indeed the
case and that for $\Lambda_s \leqsim  10^{10}$ the CCB 
constraint (at least along the $F$ and $D$ flat directions)
is merely that scalar mass squared values are positive. 

We will throughout be discussing these aspects
by assuming that there is the 
standard logarithmic running of the MSSM 
upto a scale, $\Lambda_s$, that we rather loosely 
refer to as the string scale. 
This scale may be much lower than $\Lambda_{GUT}$.
We define the QFP to be where the top Yukawa has a Landau pole at this 
point, since any variation in the Yukawa couplings above $\Lambda_s$
is expected to be drastically changed by string physics. 
As for the CCB bounds, we derive them on
the soft breaking parameters at $\Lambda_s$ since 
this is close to 
the scale at which we expect the supersymmetry breaking parameters 
to be derived in any fundamental string model (although we will have 
more to say on this in due course).

\section{The Quasi-Fixed MSSM}

The QFP~\cite{quasi,steveandme,haber} constraint, 
\ie that the top Yukawa 
coupling $h_t$ has a Landau pole at the string scale,
gives important predictions in terms of the couplings and 
masses of supersymmetric particles~\cite{quasi,steveandme,haber}.
We now examine the prediction for $h_t(m_t)$ 
numerically, paying special attention 
to its dependence on the string scale. 
Fermion masses and gauge couplings are set to be at their central values in
ref.~\cite{PDB} except for $\alpha_s(M_Z)$, which is varied to show the
induced uncertainty.
Below $m_t$, we run using a 3 loop
QCD$\otimes$1 loop QED effective theory with all superpartners integrated out.

In order to illustrate the quasi-fixed behaviour we first make a rough
calculation. To this end, we approximate the superparticle spectrum to be
degenerate at $m_t$, allowing us to use the (two-loop) MSSM renormalisation
group 
equations above that scale.
Fig.~\ref{fig:qfp1} illustrates the quasi-fixed behaviour for two values of
string
scale. The dependence of the low scale $h_t$ on its string scale value 
is shown for canonical QFP SUSY GUT framework with 
string/unification scale $\Lambda_s
=2 \times 10^{16}$ GeV. The almost horizontal part of the lines represent the
QFP regime: where, for input values $h_t(\Lambda_s) > 1.5$, 
\be h_t(m_t)=1.10 \pm 0.02 \label{pred1}\ee
results. 
Lowering $\Lambda_s$ to $10^{11}$ GeV, as in the WP
model, we see that the quasi-fixed behaviour is diminished somewhat, as
indicated by the more positive slope of the relevant lines. However, for
$h_t(\Lambda_s) > 1.5$ a QFP value of 
\be h_t(m_t)=1.17 \pm 0.04 \label{pred2} \ee
occurs\footnote{Errors quoted here include those due to the error in
$\alpha_s(M_Z)$ but they do not include those from non-degeneracy in the
superparticle spectrum.}. 

The $h_t(m_t)$ QFP prediction can be turned into a prediction of the MSSM
parameter $\tan \beta$ (the ratio of the two neutral Higgs VEVs)                   through the relation
\be
\sin \beta = \frac{\sqrt{2} m_t(m_t)}{v h_t(m_t)} \label{qfpred}
\ee
and the known value~\cite{PDB} of the top quark mass, $m_t=175 \pm 5$ GeV.
We obtain the running top mass $m_t(m_t)$ from $m_t$ by employing the 1-loop
QCD correction, thus assuming that supersymmetric corrections to it are
small.
$v$ refers to the Standard Model Higgs VEV of 246.22 GeV.
Low values of $1<\tan \beta<3$ result from
eq.~(\ref{qfpred}) when a quasi-fixed value 
$h_t(m_t)>1.05$ is used. 
The
range of $\tan \beta$ relevant here is
constrained by the non-observation of the lightest MSSM Higgs boson at
LEP2~\cite{haber}. 
The current limits~\cite{current} exclude $m_{h^0} < 107.7$ GeV for the low
$\tan \beta < 3$ scenario. 
Quasi-fixed $\tan \beta$ predictions are illustrated in table~\ref{tab:tanb},
where they are displayed
with estimated uncertainties for the WP and GUT quasi-fixed scenarios.
The uncertainties are induced by those quoted in the $h_t(m_t)$ predictions in
Eqs.~(\ref{pred1}),(\ref{pred2}).
\TABULAR[r]{c|cc}{\label{tab:tanb}
$\Lambda_s$(GeV) & $m_t=170$ GeV & $m_t=180$ GeV \\ \hline
$10^{11}$ & $1.27^{+0.13}_{-0.11}$	 &
$1.51^{+0.20}_{-0.14}$\\
$2~10^{16}$ & $1.52^{+0.10}_{-0.09} $	   & 
$1.93^{+0.08}_{-0.15}$\\
}{$\tan \beta$ prediction for a top-Yukawa QFP at the GUT scale or the WP
scale.}

\FIGURE[t]{\epsfig{file=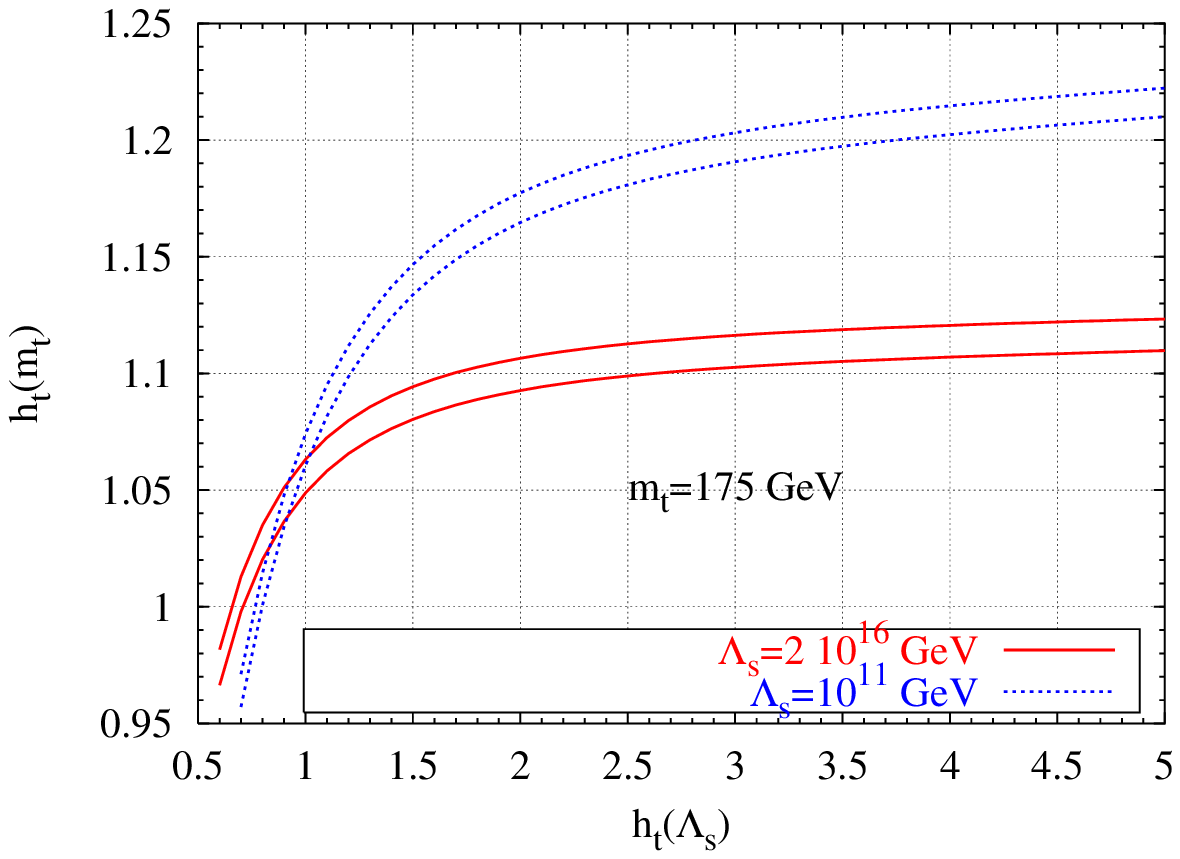, width=14cm}
\caption{\it Prediction of low energy top Yukawa coupling $h_t(m_t)$ for string
scale input $h_t(\Lambda_s)$. Two string
scales $\Lambda_s=10^{11}, 2\times10^{16}$ GeV are used. 
The pair of lines represent the range produced by varying
$\alpha_s(M_Z)=0.115-0.122$ (the upper lines corresponding to higher
$\alpha_s(M_Z)$).}
\label{fig:qfp1}
}
Here, we set $h_t(\Lambda_s)=5$,
close to its Landau pole and near the edge of perturbativity. 
In ref.~\cite{bbo}, the limit $h_t<3$ was used to define a perturbative regime
and we will use the point $h_t(\Lambda_s)=3$ as an estimator 
of sensitivity to $h_t(\Lambda_s)$.
A central value of $\alpha_s(M_Z)=0.119$~\cite{PDB} was used.
We display the results for $m_t=170,175,180$ GeV to illustrate the large
dependence upon the top mass.
We use the two-loop diagrammatic result in ref.~\cite{tanbmh} 
to calculate the MSSM lightest Higgs mass with the state-of-the-art program
{\tt
FeynHiggsFast2}.
Corrections to the values of $h_t(m_t)$ displayed in fig.~\ref{fig:qfp1} from
including sparticle thresholds are
expected to be small because the majority of change in $h_t(\mu)$ occurs in the
running between $\Lambda_s$ and 1000 GeV, identical in both cases.
We therefore use the prediction
for $h_t(m_t)$ as calculated with a degenerate sparticle spectrum at $m_t$.
To within small errors, this value should still be applicable for a
non-degenerate spectrum, which is what we assume here.

\FIGURE[t]{\epsfig{file=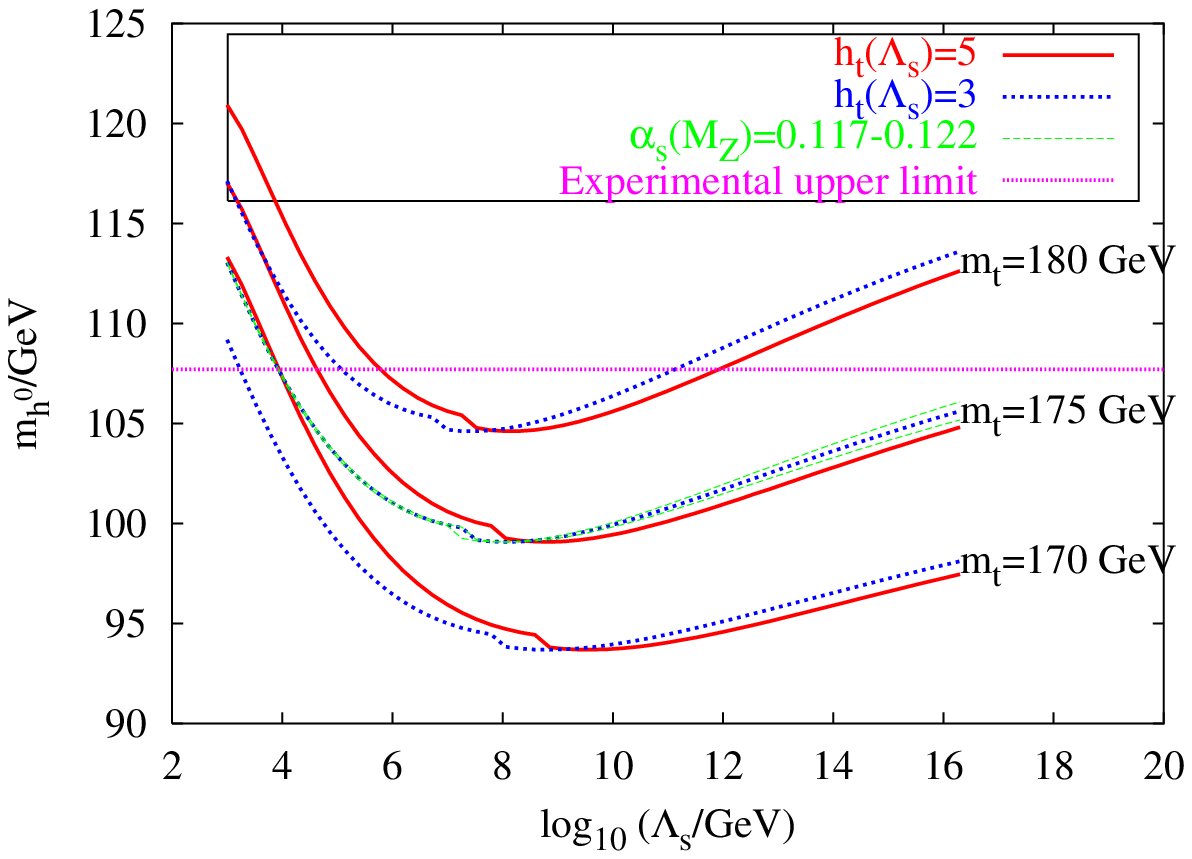, width=14cm}
\caption{\it 
Theoretical upper bound on lightest MSSM Higgs mass in the quasi-fixed
scenario with varying string 
scale $\Lambda_s$. Bounds for quasi-fixed top Yukawa couplings
$h_t(\Lambda_s)=3,5$ and $\alpha_s(M_Z)=0.119$ are shown. The copies of each
curve are for $m_t=180,175,170$
GeV from top to bottom respectively. For $m_t=175$ GeV and $h_t(\Lambda_s)=3$,
we have displayed the variation due to the error on
$\alpha_s(M_Z)=0.117-0.122$ via the lighter dashed curves.
The area underneath the experimental limit has been excluded
for the MSSM by LEP2. See text for a description of the other MSSM parameters used.
}
\label{fig:qfp2}
}

Ideally, we would now perform a parameter scan through the low energy
supersymmetry breaking parameters in order to determine the maximum value of
$m_{h^0}$ consistent with the QFP\@. This is impractical however, and we resort
to using a benchmark point in low energy supersymmetry breaking parameter
space. The value of $m_{h^0}$ obtained by the benchmark corresponds in
practice to be very close (within one GeV) to a more general upper bound on
$m_{h^0}$~\cite{tanbmh}, given an upper bound on sparticle masses.
For generality, this benchmark corresponds to {\em non-universal} SUSY
breaking parameters. 
For a given value of $\Lambda_s$, $\tan \beta$ is predicted by the QFP as in
Fig.~\ref{fig:qfp1}. We then set $\mu$ and the parameter
\begin{equation}
X_t \equiv A_t - \mu \cot \beta = 2 m_{\tilde{t}_2}. \label{cond1}
\end{equation}
As is argued in~\cite{tanbmh}, $X_t \approx 2 m_{\tilde{t}_2}$ corresponds to
the maximal-mixing case, where $m_{h^0}$ is maximised. 
$A_t$ is then specified by eq~\ref{cond1}, and therefore the gluino mass will
be set by the QFP prediction of $A_t / M_3$. For $\Lambda_s=2 \times
10^{16}$ GeV for example, we obtain $A_t / M_3=-0.59$~\cite{steveandme}.
However to the order in perturbation theory used here, the Higgs mass is
independent of the gluino mass.
Fixing $M_A$ then sets $B$ through the relation~\cite{insu}
\begin{equation}
M_A^2 = \frac{2 \mu B}{\sin 2 \beta}.
\end{equation}
The two electroweak symmetry breaking conditions are~\cite{insu}
\begin{eqnarray}
\bar{m}_1^2+\bar{m}_2^2 &=& -M_A^2, \nonumber \\
\tan^2 \beta \left( \bar{m}_2^2 + M_Z^2/2 \right) &=& \bar{m}_2^2  + M_Z^2/2,
\end{eqnarray}
where $\bar{m}_i^2 = m_{H_i}^2 + \mu^2$ plus loop corrections.
Together, they determine the
Higgs mass soft breaking parameters $m_{H_1}^2$ and $m_{H_2}^2$ (conservatively
assumed to be uncorrelated and free). 
Following the authors of ref.~\cite{tanbmh}, the maximum value of $m_{h^0}$ is
assumed 
to be acquired by taking\footnote{See ref.~\protect\cite{tanbmh} for a
definition of 
these parameters.} $M_2=100$ GeV, $M_A=1000$ GeV, $\mu=-100$ GeV and
$m_{\tilde{t}_2}=2000$ GeV in order to get a conservative estimate. 
Dependence of the upper bound on $m_{h^0}$ is logarithmic in this parameter
and therefore slowly increasing as $m_{\tilde{t}_2}$ increases. Therefore, to
obtain a sizeable effect on the bound, unnaturally high values of
$m_{\tilde{t}_2}$ would have to be taken.
Using the above procedure, the soft breaking
parameters that
$m_{h^0}$ depends most sensitively upon are fixed near the weak scale without
reference to any further unification assumptions, such as minimal supergravity
for example.

Fig.~\ref{fig:qfp2} displays the QFP value of $m_{h^0}$ predicted by the
benchmark by varying $\Lambda_s$. Uncertainties induced by the $1\sigma$ error
on $\alpha_s(M_Z)$ are shown for one particular case. It is larger for higher
$\Lambda_s$, but always less than 0.5 GeV and much smaller than the
uncertainty induced by the empirical error on $m_t$.
In fact, we see from the figure that the QFP is ruled
out to better than 1$\sigma$ for the range
\be
10^6 < \Lambda_s/\mbox{GeV} < 10^{11}
\ee
for $m_t=170-180$ GeV and $h_t(\Lambda_s)=3-5$. If we take $m_t=175$ GeV, the
QFP is ruled out for any $\Lambda_s > 10^5$ GeV. As noted above, the
$h_t(\Lambda_s)=3$ curves give an estimate of the uncertainty in the QFP
prediction. The figure shows that this dependence is small for $\Lambda_s >
10^9$ GeV but that it
increases for $\Lambda_s < 10^9$ GeV. However, we note that in this latter
range, $h_t$ being less than 5 (but still in the quasi-fixed regime)
actually {\em strengthens}\/ the upper bound upon
$m_{h^0}$. $h_t(\Lambda_s)=5$ thus gives a reasonably
accurate bound for $\Lambda_s > 10^9$ GeV and a conservative one for
$\Lambda_s < 10^9$ GeV.

\section{Analytic CCB Bounds at low string scales}

We now turn to the discussion of CCB bounds. Unphysical CCB
minima  present some of the most severe bounds 
for supersymmetric models~\cite{ccb1,dilaton,steveandme,us0,us,casas,dedes}.
Indeed, for a number of models it has been found that 
they exclude much of the parameter space not already 
excluded by experiment; for example the MSSM where
supersymmetry breaking is driven by the dilaton~\cite{dilaton}, 
SUSY GUTS at the low $\tan\beta$ quasi-fixed point (QFP)~\cite{steveandme}, 
$M$-theory in which supersymmetry breaking is driven 
by bulk moduli fields~\cite{us,casas} and
several other string/field theory scenarios~\cite{casas,dedes}. 
All of the above work, however, assumed a logarithmic evolution
of the gauge couplings with unification 
at a high scale $\geq 10^{16}$ GeV.

In this section we shall be considering the effect of 
truncating this logarithmic evolution at a low string scale.
For completeness, we first recall 
the three types of CCB minima that can occur 
in supersymmetric models:
\begin{itemize}
\item $D$-flat directions which develop a minimum due to 
       large trilinear supersymmetry breaking terms.
\item $F$ and $D$ flat directions corresponding to a single gauge invariant.
\item $F$ and $D$ flat directions which correspond to a combination 
of gauge invariants~\cite{carlos}  involving $H_2$~\cite{mango}
\end{itemize}
Since the first type are important 
at low scales~\cite{ccb1} and the second type are only 
important when there are negative mass-squared terms 
at the GUT scale, we shall concentrate on the constraints coming 
from the last type of minimum.
These occur at intermediate scales due to the running 
$H_2$ mass-squared even if all the mass-squared values are positive at the GUT 
scale. Hence the resulting constraints are very dependent 
on renormalisation group running at 
{\em high} \/scales and are particularly 
interesting from the point of view 
of models with a lower string scale.
As discussed above, our initial expectation is that the CCB bounds will be 
far less severe than in the usual versions of the MSSM.

We will consider the  
$F$ and $D$-flat direction in the MSSM corresponding to the operators
\be
L_i L_3 E_3 \mbox{~;~} H_2 L_i
\ee 
where the suffices on matter superfields are generation indices. 
With the following choice of VEVs;
\ba
\label{komkom}
h_2^0             &=& -a^2 \mu/h_{E_{33}} \nonumber \\
\tilde{e}_{L_3}=\tilde{e}_{R_3} &=& a \mu/h_{E_{33}} \nonumber \\
\tilde{\nu}_i   &=& a  \sqrt{1+a^2} \mu/h_{E_{33}}, 
\ea 
the potential along this direction
depends only on the soft supersymmetry breaking terms (neglecting a small
D-term contribution);
\be
\label{softv}
V=\frac{\mu ^2}{h_{E_{33}}^2} a^2 (a^2 (m_2^2+m_{L_{ii}}^2) + 
m_{L_{ii}}^2+m_{L_{33}}^2+m_{E_{33}}^2 ).
\ee 

In the usual MSSM we can reasonably assume that,
since the CCB minimum forms at VEVs corresponding to $a\gg 1$,
the largest relevant mass, 
and therefore the appropriate scale to evaluate the 
parameters at, 
is $\phi=h_{U_{33}} \langle h_2^0 \rangle\equiv h_t 
\langle h_2^0 \rangle $. This minimises 
the top quark contributions to the effective potential at one-loop. 
Further corrections to the potential are assumed to be small. 
Once we lower the string scale however we encounter the problem 
that the CCB minimum moves towards low scales and that consequently 
this approximation breaks down. 
Evidently, from eq.~(\ref{softv}), this happens precisely 
where the positive $m_{L_{ii}}^2 + m_{L_{33}}^2 + m_E^2$ terms begin to dominate, and 
so we do not anticipate that CCB minima will be formed when $a<1$. 
In order to check this however,  our approach will 
be to construct the constraints using the above assumption 
on $\phi$ and observe that they get far less restrictive as
we move to moderately low string scales, say $\Lambda_s \sim 10^8 \gev$.
We then check the approximate one-loop analytic results obtained with a more
accurate two-loop numerical
analysis at certain parameter points and observe numerically 
that CCB minima do not reappear as we move to very low string 
scales where $a < 1$. 

In the above potentials, $ \langle h_2^0 \rangle= -a^2 \mu /
h_{E_{33}} $ so that the eq.~(\ref{softv}) is of the form 
\be
V=\frac{\Lambda^2}{h_{U_{33}}^2} \hat\phi
 \left( \hat\phi A + B/b \right)
\ee 
where $A=m_2^2(\phi)+m_{L_{ii}}^2(\phi)$, $B$ is the $LLE$ combination 
of mass-squared parameters (also evaluated at $\phi$) that appears 
in the potential,
\be 
\hat\phi=\phi/\Lambda
\ee
and $\Lambda$ is an arbitrary scale which we shall 
take to be the usual unification scale 
$\Lambda_{GUT}\sim 10^{16}\gev$. The bound is therefore 
governed by $A$, $B$ and the parameter
\be
\label{b}
b(\phi) = \frac{\Lambda_{GUT} 
h_{E_{33}}}{h_{U_{33}} \mu}
\ee
for the $LLE,~LH_2$ direction described above, or
\be
\label{b2}
b(\phi) = \frac{\Lambda_{GUT}h_{D_{33}}}{h_{U_{33}} \mu}
\ee
for the equally dangerous $LQD,~LH_2$ direction. 

To estimate the bound, we now adapt the results of Refs.\cite{us0,us}. 
At large values of $a\gg 1$ the potential is governed by the 
first term. Whatever the string scale may be, we require that 
$m_2^2$ be positive there and negative at $M_W$
(for successful electroweak symmetry breaking). A CCB minimum radiatively
forms 
close to the value $\phi_p$ where $A$ first becomes negative 
(typically at a scale of $few\times \mu /h_{E_{33}} $)~\cite{us0,us}. 

In Refs.\cite{us0,us} it was shown that once we are able to 
estimate $\phi_p$ the bound follows fairly 
easily, and this was done for models with degenerate 
gaugino masses. Bounds were derived for 
all non-universal scalar masses and couplings.
In the present case however,
the gauge couplings and the gaugino masses 
are {\em also} \/non-degenerate at the string scale $\Lambda_s$.

This makes a general analytic treatment of the RGEs
extremely difficult, 
so in order to simplify matters we shall henceforth assume the
`GUT gaugino relation'. 
That is we assume that at the scale $\Lambda_s$ we have the 
usual GUT expression for gaugino masses, 
\be 
\label{gaugino_rel}
\frac{M_a}{M_b}=\frac{\alpha_a}{\alpha_b}.
\ee
This relationship has the useful property that 
the gaugino masses as well as the gauge couplings 
would be degenerate if we continued the evolution of the 
MSSM RGEs upto $\Lambda_{GUT}$. We shall call this fictitious 
degenerate value $M_a(\Lambda_{GUT})=M_{1/2}$. Note that
eq.~(\ref{gaugino_rel})
is only valid to one-loop order, and indeed in this section we present
analytic results to one-loop order only (contrary to the last section).

Although eq.(\ref{gaugino_rel}) may seem like a rather brutal 
requirement, it holds for a number of interesting cases,
for instance in models with power law unification as shown in 
ref.~\cite{kobayashi}. In these models the scale $\Lambda_s$ 
in our analysis should really be interpreted as the {\em compactification}\/
scale at which the 
first Kaluza-Klein states appear in the spectrum, rather than the 
string scale which is where we expect the real gauge unification to take
place after a short period of power law `running'. 
An assumption such as degenerate soft terms 
at the compactification scale $\Lambda_s$ is consistent 
with, for example, the Scherk-Schwarz mechanism of supersymmetry breaking.

Eq.(\ref{gaugino_rel}) is also expected to hold
in the mirage unification models of ref.\cite{iban}
when there is no $S/T$-mixing and in the limit 
$T+\overline{T}\rightarrow\infty$.
In this limit we have 
\be 
M_a \approx \sqrt{3}  m_{3/2} \sin\theta \frac{\alpha_a}{\alpha_0 }
+{\cal O}(1/(T+\overline{T})^2
\ee
where we use the subscript-0 to represent values at the 
usual $\Lambda_{GUT}$ unification scale (\ie $\alpha_0 \approx 1/25 $), and
where we have neglected terms of order $\alpha_a m_{3/2} $ which 
is consistent to one-loop accuracy. 
In this case we have $M_{1/2} = \sqrt{3}  m_{3/2} \sin\theta$.

Eq.(\ref{gaugino_rel}) allows us to adapt the expressions 
of ref.\cite{us0} with only a modest amount of effort
by writing the parameters at $\Lambda_s$ in terms of their 
values at $\Lambda_{GUT}$. 
In order to proceed, we next spend a little time discussing the 
analytic solutions to the renormalisation group running.
The solutions of all the parameters may easily be expressed in terms of 
those combinations with infra-red QFPs; 
$R=h_t^2/g_3^2$, $A_t$ and $3 M^2
= m_2^2+m^2_{U_{33}}+m_{Q_{33}}^2$.
These may be written as functions of 
\be
r = \frac{\alpha_0}{\alpha_3} \equiv \frac{1}{\tilde{\alpha}_3}=
1+\frac{6\alpha_0}{4\pi}\log \frac{\Lambda}{\Lambda_{GUT}},
\ee
so that
\be
\frac{\alpha_0}{\alpha_2} \equiv \frac{1}{\tilde{\alpha}_2}=\frac{3}{4-r}
\mbox{~;~} 
\frac{\alpha_0}{\alpha_1} \equiv \frac{1}{\tilde{\alpha}_1}=
\frac{5}{16-11 r}.
\ee
Taking $\alpha_3(m_t)=0.108$ means that $0.37<r<1$
with $r=1$ corresponding to the GUT scale.
If the string scale is at $\Lambda_s=10^{11}\gev $ as in the WP model, then 
the corresponding value of $r_s\equiv r(\Lambda_s)$ is $r_s= 0.82$.
It is useful to define 
\ba
\Pi(r) &=& \tilde{\alpha}_3^{16/9}
\tilde{\alpha}_2^{-3}
\tilde{\alpha}_1^{-13/99}
\nn\\
\hat{J}&=& \frac{1}{r\Pi(r)}\int ^1_r \Pi(r') \dif r'.
\ea
Solving for $R$ in terms of its value $R_s$ at the 
string scale (we use subscript-$s$ to denote string-scale values) we find 
\be
\frac{1}{R} = \frac{\Pi_s r_s }{R_s \Pi r }
+ \frac{1}{R^{QFP}}
\ee
where the QFP value (where the Yukawa couplings blow up at the 
string scale) is given by 
\be
\label{Rqfp}
\frac{1}{R^{QFP}}=
2 \hat{J}(r) - 2 \hat{J}(r_s) \frac{\Pi_s r_s}{\Pi r}. 
\ee
We also, for later use, define the distance from the 
real QFP, 
\be
\sigma = \frac{R}{R^{QFP} }.
\ee
This can be rewritten in terms of a fictitious renormalisation 
of $R$ down from a $\Lambda_{GUT}$ scale value of $R_0$;
\ie defining  
\be
\label{Rqfpbar}
\frac{1}{\overline{R}^{QFP}}= 2\hat{J}
\ee
we have 
\ba
\frac{1}{R} &=& \frac{1}{R_0 \Pi r }
+\frac{1}{\overline{R}^{QFP}} \nn\\
\frac{1}{R_s} &=& \frac{1}{R_0 \Pi_s r_s }
+\frac{1}{\overline{R}_s^{QFP}}.
\ea
This is the usual expression for $R$ (\cf ref.\cite{us}); however it should 
be noted that $R_0$ is here merely a parameter that is 
negative in the region 
$1/\overline{R}^{QFP} > 1/R_s > 0 $. In the usual MSSM with unification 
at the GUT scale, this would of course be an unphysical (non-perturbative)
region.
For $A_t$ and $M^2$ we now define the distance from the 
usual QFP (\ie where couplings blow up at the usual unification 
scale $\Lambda_{GUT}$) 
\be
\rho = \frac{R}{\overline{R}^{QFP}}
\ee
and also
\be
\xi = \frac{1-r}{r\hat{J}} -1.
\ee
We then obtain expressions for $\tilde{A}_t=A_t/M_{1/2}$ and  
$\tilde{M}^2 = M^2/M_{1/2}^2 $ in terms of their fictitious values, 
$\tilde{A}_0$ and $\tilde{M}^2_0$, at $\Lambda_{GUT}$;
\ba 
\tilde{A}_t &=& (1-\rho) \tilde{A}_0
+  \rho \xi - \Gamma  \nn\\
\tilde{M}^2 &=&  (1-\rho ) \tilde{M}^2_0
- \frac{1}{3}\rho \tilde{K} +\frac{2}{3} (1-r)\gamma
\ea
where 
\ba
\gamma &=& \frac{16}{9}\tilde{\alpha}_3 (1+\tilde{\alpha}_3 (1-r)/2)
+\tilde{\alpha}_2(1-\tilde{\alpha}_2 (1-r)/6) 
+\frac{13}{45} \tilde{\alpha}_1(1-11\tilde{\alpha}_1 (1-r)/10)\nn\\
\Gamma &=& (1-r) \left(
\frac{16}{9}\tilde{\alpha}_3 +
\tilde{\alpha}_2 +
\frac{13}{45} \tilde{\alpha}_1\right) \nn \\
\tilde{K} &=& (1-\rho)
(\xi -\tilde{A}_0)^2 - \xi^2 + (\xi+1)\Gamma
\nn\\
\tilde{A}_0 &=& \left( \tilde{A}_s -\rho_s\xi_s + \Gamma_s\right) / 
(1-\rho_s) \nn\\
\tilde{M}^2_0 & = & \left( \tilde{M}_s^2 
+ \frac{1}{3}\rho_s \tilde{K}_s -\frac{2}{3}(1-r_s)\gamma_s \right) 
/(1-\rho_s).
\ea
It is important to note that, since 
\be
1-\rho = (1-\sigma ) (1-\rho_s ) , 
\ee
$A_t$ and $M^2$ retain their QFP
behaviour since when $\sigma = 1 $ (or $R_s\rightarrow \infty $)
they are both independent of their 
values {\em at the string scale}, $\Lambda_s$.
In addition, factors of $1/(1-\rho_s)$ cancel so that there 
is no divergent behaviour at the usual QFP\@.
Also note that this QFP is at lower $\tan\beta $ than in the 
usual MSSM unification. We can estimate the difference 
in $\tan\beta$ at the QFP by using 
\be 
R=\frac{m_t^2 }{4\pi \alpha_3 v^2 \sin^2\beta },
\ee
so that 
\be 
\sin^2 \beta^{QFP} = \frac{\overline{R}^{QFP}}{R^{QFP}} 
\sin^2 \overline{\beta}^{QFP}.
\ee
Eqs.~(\ref{Rqfp},\ref{Rqfpbar})
then give 
$\tan\beta^{QFP}\approx 1.2$
in the WP model with $\Lambda_s=10^{11}\gev $, in agreement with the full
two-loop numerical result presented in 
fig.~\ref{fig:qfp1}. 

With all parameters expressed in terms of 
GUT scale parameters, we are now simply able to apply the 
bounds derived in ref.\cite{us} for non-universal SUSY breaking
directly. 
Consider for example the $LH_2$, $LLE$ direction. The cosmological bounds 
in this case are 
\be
\label{arp4}
\left.(2 \tilde{m}_{L_{ii}}^2 + \tilde{m}_2^2-\tilde{m}^2_{U_{33}}-
\tilde{m}^2_{Q_{33}})\right|_0
\geqsim 
f(\tilde{B}|_0)+(\rho_p-1)\left( g(\tilde{B}|_0)+3 \tilde{M}^2|_0
-\rho_p (1-\tilde{A}_0)^2 \right),
\ee
where $\rho_p $ is the value of $\rho $ at the scale $\phi_p$
and 
\ba
f(x) &=& 1.20-0.14 x + 0.02 x^2 \nn\\
g(x) &=& 2.77-0.18 x + 0.02 x^2 \nn\\
B &=& m^2_{L_{ii}}+m^2_{L_{33}}+ m^2_{E_{33}},
\ea
for $\mu = 500 \gev$. (The small dependence of $f$ and $g$ on $\mu$, 
which we must choose by hand, is discussed in ref.\cite{us}.)
To a good approximation the 
value of $\rho_p$ is given by~\cite{us}
\be 
\frac{1}{\rho_p} = 
1+\frac{1}{2 R_0} = 1+3.17 (\sin^2\beta -\sin^2 \overline{\beta}^{QFP}).
\ee
In order to relate the quantities to their string scale values, 
we use the one loop RGE solutions for $A$ and $B$; 
\ba
\label{arp5}
\left.(2 \tilde{m}_{L_{ii}}^2 + \tilde{m}_2^2-\tilde{m}^2_{U_{33}}-
\tilde{m}^2_{Q_{33}})\right|_s
& \geqsim & -\frac{16}{9}\delta^{(2)}_{3s} - 3 \delta^{(2)}_{2s} - 
\frac{5}{99} \delta^{(2)}_{1s} +  
f(\tilde{B}|_0) \nn\\
&&+(\sigma_p-1)\left\{ (1-\rho_s)  g(\tilde{B}|_0)+3 \tilde{M}^2|_s
- 2 (1-r_s)\gamma_s  \right. \nn\\
&&
+\rho_s \left(-1 + \rho_s(\xi_s-1)^2 - \Gamma_s (\xi-3)
 - 2\tilde{A}_s (\xi-1)\right) \nn\\
&&\left. -\sigma_p (\tilde{A}_s - \rho_s\xi_s + \Gamma_s + \rho_s-1 )^2
\right\},
\ea
where 
\ba 
\delta^{(n)}_i &=& \frac{\alpha_i^n}{\alpha^n_0}-1 \nn\\
\left.\tilde{B}\right|_s &=& \left.\tilde{B}\right|_0 - 3 \delta_{2s}^{(2)}
-\frac{1}{11}\delta^{(2)}_{1s},
\ea
and where 
\be 
\sigma_p = 1 - \frac{\Pi_sr_x}{\Pi_sr_s (1-\rho_s ) + 2 R_s}.
\ee

\FIGURE[t]{\epsfxsize=14cm
\epsffile{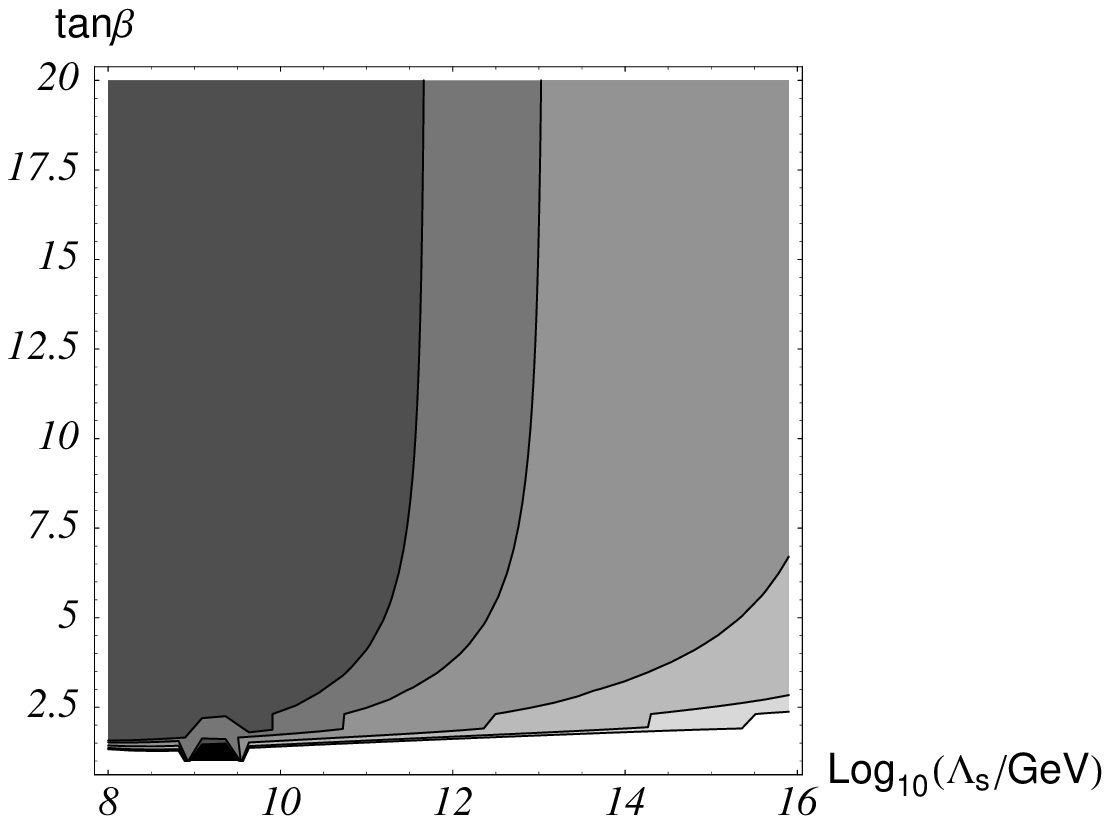}
\caption{\it Charge and colour breaking bounds 
with a lower string scale, $\Lambda_s$,
for $\mu=500\gev$ and degenerate trilinear terms, 
$A=-M_{1/2}$, and scalar masses $m_s$, at the string scale. 
The figure shows bounds on 
$\tilde{m}_s^2= m_s^2/M_{1/2}^2 $ for 
varying $\Lambda_s$ and $\tan\beta$ (i.e.\ away from the QFP). 
The contours are
$\tilde{m}_s^2> 0$ (black), $\tilde{m}_s^2> 0.25$ (medium dark),
$\tilde{m}_s^2> 0.33$ (medium), $\tilde{m}_s^2> 0.5$ (medium light), 
$\tilde{m}_s^2> 0.66$ (light), $\tilde{m}_s^2> 0.75$ (white). } 
\label{fig:fig1}}
The general behaviour of the bounds is clearly similar to that 
in the usual unification scenario. The bounds are on the 
particular combination
$\left.(2 \tilde{m}_{L_{ii}}^2 + \tilde{m}_2^2-\tilde{m}^2_{U_{33}}-
\tilde{m}^2_{Q_{33}})\right|_s$ and are most restrictive at the QFP, 
decreasing as $\tan\beta $
increases. Away from the QFP there 
is a quadratic dependence on $\tilde{A}_s$ with 
a minimum at $\tilde{A}_s = {\cal{O}}(1)$. 

We can now see why the bounds at low scales 
are far less severe than in the MSSM with unification 
at the GUT scale. First, close to the QFP, the bound is 
\ba
\label{arp6}
\left.(2 \tilde{m}_{L_{ii}}^2 + \tilde{m}_2^2-\tilde{m}^2_{U_{33}}-
\tilde{m}^2_{Q_{33}})\right|_s
& \geqsim & -\frac{16}{9}\delta^{(2)}_{3s} - 3 \delta^{(2)}_{2s} - 
\frac{5}{99} \delta^{(2)}_{1s} + f(\tilde{B}|_0) \nn\\
&=&  -0.48 + f(\tilde{B}|_0),
\ea
for $\Lambda_s=10^{11}\gev $.
Thus the non-degeneracy of gauge couplings and gauginos 
contributes negatively to the bound even at the QFP\@. Second, 
away from the QFP, the bound asymptotes to the values with 
\be 
\label{ltanb}
\rho_p = \frac{1}{1+3.17 \cos^2 \overline{\beta}^{QFP}} \sim 0.57.
\ee
However, the quantity multiplying $\tilde{M}^2_s$ in the bound 
is now $(\sigma_p-1)$ which is a larger {\em negative} \/factor than 
$(\rho_p-1)$.

\FIGURE[t]{\epsfig{file=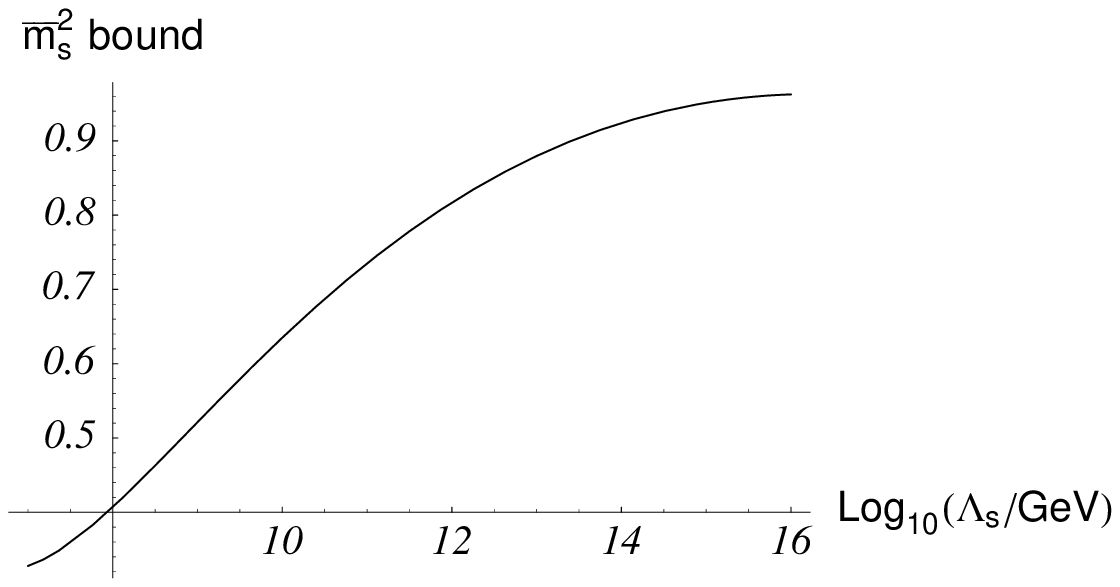, width=14cm}
\caption{\it Charge and colour breaking bounds 
with a lower string scale at the quasi fixed point (QFP)
for $\mu=500\gev$. 
The figure shows lower bounds on 
the string scale values of $\tilde{m}_s^2= m_s^2/M_{1/2}^2 $
for varying $\Lambda_s$.} 
\label{fig:fig2}}
We now further specialise to the mirage unification models with $V_0=0$, 
which have degenerate $A$-terms and degenerate scalar masses at the string 
scale;
\ba
\label{mirage}
\tilde{A}_s &=& -1 \nn\\
\tilde{m}_{s}^2 &=& \mbox{unconstrained}.
\ea
Contours of the $LH_2$, $LLE$ bound are shown in fig.~\ref{fig:fig1}, 
for varying $\tan\beta$ and $\Lambda_s$. 
The diagram shows that a lower string
scale removes the dangerous minima. 
Indeed, for 
the WP model value of $\Lambda_s \sim 10^{11}\gev$,
there are {\em no} \/CCB minima 
appearing along the $LH_2$, $LLE$ direction 
except close to the QFP ($\tan\beta\leqsim 3$)
or for negative scalar mass squared values ($m_s^2<0$). At the QFP we find that
the bound at 
$\Lambda_s=2~10^{16}$ GeV is $\tilde{m}_s^2 
\geqsim 0.95$ but drops rapidly towards smaller 
values of $\Lambda_s$, as shown in fig.~\ref{fig:fig2}. 
A full numerical determination of the bounds for specific points in 
parameter space is in accord with Figs.~\ref{fig:fig1} and~\ref{fig:fig2}. It
also shows that the bounds are in fact not 
overly sensitive to the precise values of 
$\alpha_1$ and $\alpha_2$ at $\Lambda_s$ since
the running is dominated by $\alpha_3$.

Moreover, this behaviour is expected to be a general feature
resulting from the low string scale pushing the CCB minimum 
to low scales. For example we can analyse the bound at large $\tan\beta $ 
where eq.~(\ref{ltanb}) holds. Choosing $M_s^2=0$ and adjusting 
$A_s$ to make $A_0=M_{1/2}$, one finds that, away from the QFP,  
there are no CCB minima for any positive choice of non-universal 
mass-squared parameters at the string scale for 
$\Lambda_s \leqsim 10^{10}\gev $. In other words, for these intermediate 
and low string scales one may always adjust $A_s$ to remove CCB minima.
Conversely, choosing a large enough value of $A_s$ forms a CCB minimum at 
any $\Lambda_s$.

For $\Lambda_s \leqsim 10^7\gev$ the analytic approximations we have been 
using break down for reasons outline above. Specifically, instead of 
evaluating the parameters at the renormalisation scale $\phi=h_t \langle
h_2^0 \rangle $, it is now more accurate to evaluate them at
the scale $\phi = g_2 \langle l \rangle $ (in the $LLE,LH_2$ direction) 
since this would be the largest relevant mass. 
Using this definition for $\phi$ we 
find numerically that minima do not reappear when 
$\Lambda_s$ is lowered still further, as 
expected due to the dominance of the positive 
$m_{L_{ii}}^2+m_{L_{33}}^2+m_E^2$ contribution to the potential at low VEVs. 

\section{Summary}

To summarise, we have examined constraints on the MSSM coming from the QFP
scenario and CCB bounds when the string scale is lower than the
canonical unification value of $10^{16-17}$ GeV.
The quasi-fixed behaviour is weakened somewhat as the scale is reduced, 
\ie weak MSSM parameters retain more information about 
their high energy boundary
conditions. Very strict bounds upon the string scale are obtained from
the LEP2 lower bound upon the lightest Higgs mass in the QFP scenario.
Current limits exclude the QFP scenario for string scales 
between $10^6$ and $10^{11}$ GeV for $m_t=175 \pm 5$ GeV.
This range of exclusion will increase by the end of running of LEP2, as the
bounds improve. 
Run II of the Tevatron is expected to decrease the errors upon $m_t$
significantly, with important implications for the range of $\Lambda_s$ ruled
out in the quasi-fixed scenario. For example, an error of 1 GeV upon $m_t$
would rule out the QFP scenario for all $\Lambda_s > 10^5$ GeV.

CCB bounds also give important constraints upon the quasi-fixed scenario.
We provided an analytic treatment of CCB bounds with 
lower string scales which we confirmed with a more accurate numerical check. 
It is clear from our results that lowering the string scale
significantly weakens the CCB bounds. As an example, we considered the most
restrictive case of the QFP\@. In this case the lower bound upon 
string-scale, degenerate, scalar mass-squared values $\tilde{m}_s^2$
is weakened by 30\% in the WP model, $\Lambda_s=10^{11}$ GeV. Remarkably, for
$\tan \beta>2$ and $\Lambda_s < 10^{10}$ 
GeV, the CCB bound is merely $m^2 > 0$ for {\em any}\/ non-universal pattern 
of supersymmetry breaking. 

Although we have concentrated on a particular subset of models (\ie
those that preserve the `GUT gaugino relation'), we argue that our conclusions
are true
in a more general case. As the string scale is lowered, provided that all
mass-squared values are initially positive, the CCB minima are inevitably
pushed to lower VEVs.  At these low scales, the negative $m_2^2 $ term
no longer dominates the potential along the most dangerous $F$ and
$D$-flat directions.

\acknowledgments
This work was partially supported by PPARC\@. We would like to thank 
T. Gherghetta, L. Ibanez and F. Quevedo for helpful discussions. 
BCA would like to thank the CERN theory division (where this work 
was initiated) for hospitality offered.

\end{document}